\documentclass[twocolumn,prl,showpacs,floatfix]{revtex4}
\usepackage{graphicx}

\usepackage{bm}
\begin{document}

\title{Mechanical restriction versus human overreaction triggering congested traffic states}

\author {Hyun Keun Lee,$^1$ Robert Barlovic,$^2$ Michael Schreckenberg,$^2$ and
Doochul Kim$^1$} \affiliation{$^1$School of Physics, Seoul
National University, Seoul 151-747, Korea }
\affiliation{$^2$Theoretische Physik Fakult\"at 4, Universit\"at
Duisburg-Essen, D-47048 Duisburg, Germany }
\date{\today}

\begin{abstract}

A new cellular automaton (CA) traffic model is presented. The
focus is on mechanical restrictions of vehicles realized by
limited acceleration and deceleration capabilities. These features
are incorporated into the model in order to construct the
condition of collision-free movement. The strict collision-free
criterion imposed by the mechanical restrictions is softened in
certain traffic situations, reflecting human overreaction. It is
shown that the present model reliably reproduces most empirical
findings including synchronized flow, the so-called {\it pinch
effect}, and the time-headway distribution of free flow. The
findings suggest that many free flow phenomena can be attributed
to the platoon formation of vehicles ({\it platoon effect}).

\end{abstract}
\pacs{ 89.40.-a, 45.70.Vn, 05.45.-a, 05.20.Dd
} \maketitle


Traffic flow phenomena have been analyzed and modelled from the
viewpoint of statistical physics since the early 1990s. The main
issues are the characterization of traffic phases and transitions
among them as well as the development of appropriate traffic
models. The investigation of empirical data has lead to the
identification of three traffic phases (free, synchronized,
jammed)~\cite{tgf03kerner,HeRe,ChRe,Hys,3Ph,3PhTr,EmpR}.
Furthermore, time-headway distributions extracted from single
vehicle data have been reported~\cite{TiHe}, showing that even
free flow is not as trivial as previously believed. In order to
reproduce the empirical findings, many traffic models have been
proposed both from the microscopic~\cite{NSCA,BaOp,KrLd} and the
macroscopic~\cite{KeFl} viewpoints. Also, there are efforts to
establish a link between them~\cite{BeLi,HKLi,HeLi,HKLi2} in order
to probe any universal features. These achievements have deeply
influenced the understanding of traffic flow far from equilibrium
with interacting particles showing collective behaviors.

A primary objective of traffic models is to explain the
synchronized flow phase that is characterized by a considerably
high flux without any clear density-flux relation. Unfortunately,
most models produce merely the free flow and jammed traffic. The
fluctuating profiles, appearing during the transient process
heading towards the phase separation of the two phases above, are
regarded as the synchronized flow in those models. However, such a
behavior is not compatible with the stability of the synchronized
flow found in empirical data. The approach of Knospe {\it et
al.}~\cite{KnBL} produces a different type of solution for the
synchronized flow, which is not a part of any transient process.
However, the single vehicle data show unrealistic velocity
fluctuations in this solution. Recently, Kerner {\it et al.} have
claimed to be able to reproduce the synchronized flow~\cite{KeCL}.
However, in their model, a phenomenological requirement for
synchronized flow is directly implemented, namely the driver's
attempt to adopt the velocity of the vehicle in front within the
so-called synchronization distance. Another shortcoming of
existing traffic models is that even the free flow is not
reproduced so successfully. In~\cite{TiHe}, it is reported that a
small time-headway below $1$ sec is frequently observed especially
in the free flow. Simple tuning of the spatiotemporal scale to
achieve such a small time-headway results in an extraordinarily
high flux. A method to bypass this problem is to utilize the gap
between the next two vehicles in front~\cite{KnBL}, leading to a
time-headway definitely below $1$ sec. However, the coupling of
two vehicles is inevitable in this approach, which is not observed
empirically~\cite{NoCo}. In this Letter, a new {\it Cellular
Automaton} (CA) traffic model is presented, which overcomes the
problems stated above. The mechanical restriction is introduced in
the form of limited acceleration and deceleration capabilities.
Also, human behavior is modelled as the driver's excessive
response to the local traffic conditions. The former is a natural
consequence of given physical limitations while the latter is
incorporated into the model to reflect human overreaction. In the
following, all variables are assumed to be integer numbers unless
stated otherwise.

First, we introduce the limited capability of acceleration ($a$)
and deceleration ($D$). For simplicity, these are both assumed to
be constant. The idea of limited acceleration is implemented in
most existing CA traffic models. However, the deceleration
limitation has not been enforced strictly or systematically. It is
important to note that the bounded braking capability changes the
collision-free mechanism entirely. Most CA traffic models impose a
collision-free condition explicitly by assigning arbitrary
deceleration values required to prevent collision. However, it is
rather natural to view the collision-free flow as a consequence of
moderate driving instead of infinite braking capabilities. For
such a physically realizable flow, we first design a heuristic
collision-free driving dynamics strictly observing the limited
deceleration. The starting point of the new CA model is an
inequality which guarantees safe driving. Here a vehicle prepares
for the worst case, namely that the leading vehicle may brake
suddenly at any time $t$. Since the follower's reaction is delayed
due to the response time which is assumed to be the unit time of
the model, whether secure driving is possible or not is determined
at time $t+1$ when the reaction begins. The velocity which allows
safe movement, represented by $c_n^{t+1}$ below, should satisfy
\begin{equation}
   x_n^t + \Delta + \sum_{i=0}^{\tau_{\rm
   f}(c_n^{t+1})}(c_n^{t+1}-Di)
   \le x_{n+1}^t + \sum_{i=1}^{\tau_{\rm l}(v_{n+1}^t)}(v_{n+1}^t-Di),
 \label{dcf}
\end{equation}
where $x_n^t$ ($v_n^t$) is the location (velocity) of the $n$-th
vehicle at time $t$ and the increased index represents the vehicle
in front. $\Delta$ is the minimal coordinate difference required
by the follower to guarantee its safety, and thus assumed to be at
least the length of a vehicle $L$. Each summation accounts for
successive decelerations during time steps $i=0,1,..,\tau_{\rm f}$
($i=1,..,\tau_{\rm l}$) with maximum braking capability $D$, where
$\tau_{\rm f}$ ($\tau_{\rm l}$) for the follower (leader) will be
specified below. The zero-based summation index stands for the
response time of the follower. For $\tau_{\rm f,l}(v)=v/D$ and
$\Delta = L$, Eq.~(\ref{dcf}) suggests such $c_n^{t+1}$ that can
guarantee a complete stop showing bumper-to-bumper configuration.
We call this dynamics the strict collision-free dynamics (or
criterion) in this work. Later on, the expressions for $\tau_{\rm
f,l}$ and $\Delta$ will be modified to take into account the {\it
human overreaction}. Note that for a given $n$, the safe velocity
$c_n^{t+1}$ is not unique but just has an upper bound. In the
following, the largest $c_n^{t+1}$ satisfying Eq.~(\ref{dcf}),
denoted by $\tilde{c}_n^{t+1}$, is used to reflect the desire of
drivers to move as fast as possible.

Next, an element of human behavior is introduced which is actually
responsive to the local traffic situation. Generally, it is
accepted that a driver's maneuvers are not precisely predictable
by a simple rule. The usual solution for this problem in
simulation models is to introduce fluctuations which cover the
human factor in a stochastic way. However, a different strategy is
adopted here. It is supposed that the driver's behavior may be
biased depending on the local traffic situation. To implement this
simply, a {\it 2-state} variable is introduced:
\begin{equation}
\gamma_n^t = \left\{
   \begin{array}{ll}
      0 &~~{\rm for}~v_n^t \le v_{n+1}^t \le v_{n+2}^t~{\rm or}
                    ~v_{n+2}^t \ge v_{\rm fast}, \nonumber \\
      1 &~~{\rm otherwise} \nonumber
   \end{array}
\right.
\label{gamma}
\end{equation}
with a constant $v_{\rm fast}$ slightly below $v_{\rm max}$. The
state $\gamma_n^t=0$ corresponds to a situation where the driver
judges that the local situation is optimistic since the cars in
front are speeding away. It is assumed that the driver will move
faster in order to catch the car ahead, even faster than the
allowed velocity under the strict collision-free dynamics. This
state is denoted as the {\it optimistic state}. Otherwise, for
$\gamma_n^t=1$, the driver is in the {\it defensive state}, where
the vehicle slows down below the velocity due to the strict
collision-free criterion. This distinction in respect to the local
traffic situation is termed {\it human overreaction} here. It is
realized in the model by manipulating $\tau_{\rm f,l}(v)$ and
$\Delta$ in (\ref{dcf}) as follows:
\begin{equation}
\begin{array} {ll}
   ~\Delta &= L + \gamma_n^t
             {\rm max}\{0,{\rm min}\{g_{\rm add},v_n^t-g_{\rm add}\}\}
               , \nonumber \\
   \tau_{\rm f}(v) &= \gamma_n^t v/D + (1-\gamma_n^t){\rm max}\{0,{\rm min}\{v/D,t_{\rm
   safe}\}-1\}, \nonumber \\
   \tau_{\rm l}(v) &= \gamma_n^t v/D + (1-\gamma_n^t){\rm min}\{v/D,t_{\rm safe}\}
               . \nonumber
\end{array}
\label{modtd}
\end{equation}
Herein $g_{\rm add}$ is introduced for an additional security gap
in the {\it defensive state} ($\gamma_n^t = 1$) and $t_{\rm safe}$
is a maximal time step during which the follower observes his/her
own safety in the {\it optimistic state}. The additional $-1$ for
$\tau_{\rm f}(v)$ compensates for the surplus time step due to the
follower's response time only when $\gamma_n^t = 0$, and thus the
role of $t_{\rm safe}$ is properly implemented. For
$\gamma_n^t=0$, the $\tau_{\rm f,l}$ can be smaller than those
($v/D$) necessary for complete stops while $\Delta$ returns to
$L$. Consequently, lower safety is required compared to that
needed by strict collision-free dynamics, and thus a faster
$\tilde{c}_n^{t+1}$ can be chosen. On the other hand, for
$\gamma_n^t=1$, $\Delta$ can be larger than $L$ while $\tau_{\rm
f,l}(v)$ return to $v/D$, which implies over safety. In this way,
a lower $\tilde{c}_n^{t+1}$ can be assigned.

The update rules of the model can be written in the following
form:
\newline
$~~~~1~p = {\rm max}\{p_{\rm d},p_0-v_n^t(p_0-p_{\rm d})/v_{\rm
slow}\}$
\newline
$~~~~2~\tilde{c}_n^{t+1} = {\rm
max}\{c_n^{t+1}|~c_n^{t+1}~\mbox{satisfies
Eqs.~(\ref{dcf}--\ref{modtd})}\}$
\newline
$~~~~3~\tilde{v}_n^{t+1} = {\rm min}\{v_{\rm max},v_n^t+a,
                      {\rm max}\{0,v_n^t-D,\tilde{c}_n^{t+1}\}\}$
\newline
$~~~~4~v_n^{t+1} = {\rm max}\{0,v_n^t-D,\tilde{v}_n^{t+1}-\eta\}$
\newline
$~~~~~~~{\rm where}~\eta = 1~{\rm if~rand()}<p,~{\rm or}~0~{\rm
otherwise}$
\newline
$~~~~5~x_n^{t+1} = x_n^t+v_n^{t+1}.$
\newline
Herein, the stochastic parameter $p~(<1)$ in step 1 linearly
interpolates between $p_0$ and $p_{\rm d}$ if $v_n^t$ is smaller
than $v_{\rm slow}$ (note that $p$ is a real number). We set
$p_0>p_{\rm d}$ so that step 1 is a generalization of the well
known {\it slow-to-start} rule~\cite{KnBL,Barlo1,Barlo2}, which is
known to be an ingredient in the formation of congested traffic
states. Step 3 guarantees that the updating velocity satisfies the
mechanical restriction as well as the traffic regulation. The
stochastic deceleration in step 4 is also limited by the braking
capability $D$. Note that steps 3 and 4 observe the limited
deceleration consistently~\cite{Comment1}. The length of one cell
is chosen to be $\Delta x=1.5$ m and the unit time is set to
$\Delta t = 1$ sec. The following model parameters are motivated
by empirical facts: $a=1$, $D=2$, $L=5$, $v_{\rm fast}=19$,
$t_{\rm safe}=3$, $g_{\rm add}=4$, $p_0=0.32$, $p_{\rm d}=0.11$,
$v_{\rm slow}=5$, and $v_{\rm max}=20$.

\begin{figure}
\includegraphics[width=8.0cm, height=6.0cm]{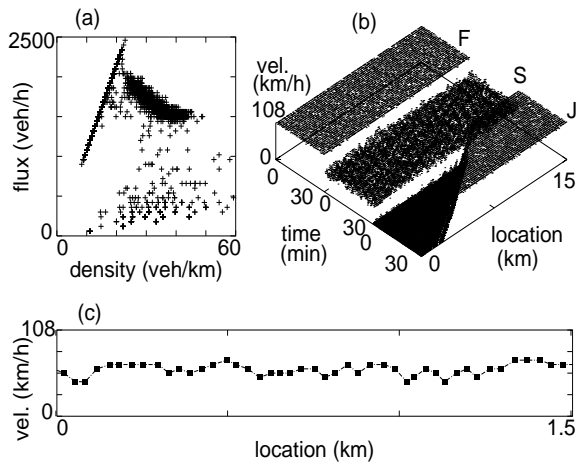}
\caption{Results for a road (periodic boundary conditions)
consisting of $40,000$ cells ($60$ km). A homogeneous distribution
of standing vehicles is used as the initial condition while the
density is varied from 10--50 veh/km with a step of 2 veh/km. (a)
Fundamental diagram: Single vehicle data is gathered at a fixed
position and then the flux ($J$) and velocity ($v$) are averaged
every minute. The density ($\rho$) is simply obtained via the
hydrodynamic relation $J = \rho v$. The measuring time is $10,000$
sec after the relaxation of $30,000$ sec. (b) The spatiotemporal
shape of the different traffic phases [F(ree), S(ynchronized),
J(ammed)] is depicted for $30$ min. The initial densities are
$16$, $30$, and $44$ veh/km, respectively. (c) Snapshot of
synchronized flow (randomly chosen part zoomed at $40,000$ sec)
with an initial density of $30$ veh/km. The filled squares
represent vehicles moving from left to right.} \label{f.fund}
\end{figure}
In Fig.~\ref{f.fund} the traffic phases occurring under periodic
boundary conditions are depicted. Three traffic states (free flow,
synchronized traffic, jams) can be identified in the fundamental
diagram Fig.~\ref{f.fund}(a) (see~\cite{HeRe}). The straight line
with the positive slope corresponds to free flow. The synchronized
states form a 2-dimensional region in the middle of the diagram
while jammed vehicles produce the scattered points below. A
typical spatiotemporal shape of each phase is shown in
Fig.~\ref{f.fund}(b). Note that the so-called {\em universal}
constants of traffic flow~\cite{Hys,3Ph,3PhTr}, such as the jam
velocity $v_{\rm g}\approx -15$ km/h and the flux out of a jam
$q_{\rm out}\approx 1800$ veh/h, are also reproduced by simply
adjusting $p_0$ and $p_{\rm d}$. For the purpose of demonstrating
the inner structure of synchronized flow a snapshot of a part of
the road is presented in Fig.~\ref{f.fund}(c). As required, this
is not a transient process and exhibits a smooth velocity profile.
This implies that the points comprising the synchronized area in
Fig.~\ref{f.fund}(a) are not attributed to the averaged effects of
strong fluctuations but to the special headway-velocity relation.

\begin{figure}
\includegraphics[width=8.0cm, height=9.0cm]{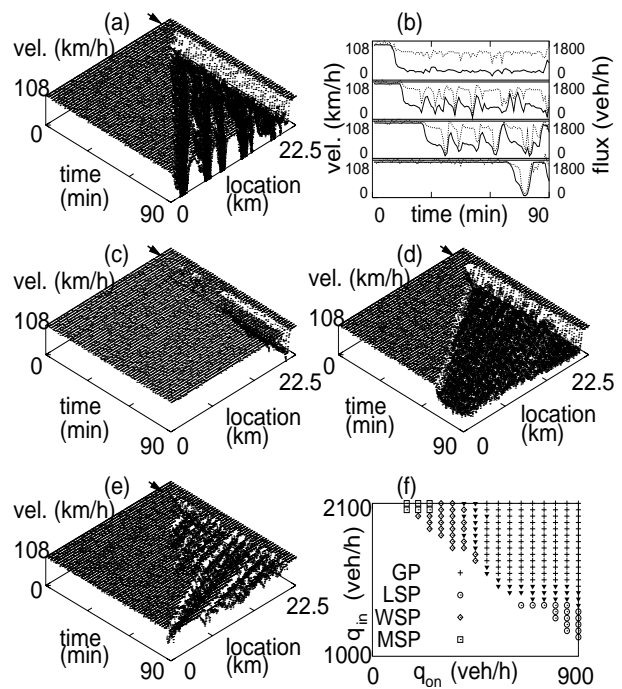}
\caption{ The impact of an on-ramp at $20.4$ km (black arrow) with
a $0.6$-km long merging road is demonstrated. The left (right) end
of the road is positioned at $-45$ km ($45$ km). The influx at the
left is kept constant during simulation. After $q_{\rm in}$ is
supplied for $10$ hours, the simulation time is set to $0$ and
then $8$ minutes later the on-ramp injection $q_{\rm on}$ is
turned on. Within the merging road the widest leading gap is
selected and then a vehicle is inserted at the mid point. For its
velocity, $0.7$ times that of the follower's is assigned. Below, a
coupled number means $(q_{\rm in},q_{\rm on})$. (a) General
pattern (GP) at $(1800,550)$. (b) Pinch effect in (a): Each figure
shows one minute averaged velocity (solid line) and flux (dotted
line) at different locations. From the top, detectors are located
at $20.4$, $18.9$, $15.9$, and $5.4$ km, respectively. (c)
Localized synchronized pattern (LSP) flow at $(1300,650)$. (d)
Widening synchronized pattern (WSP) flow at $(1950,350)$. (e)
Moving synchronized pattern (MSP) flow at $(2050,150)$. (f) Phase
diagram of the congested traffic patterns. The filled triangles
correspond to mixed patterns.} \label{f.ramp}
\end{figure}
The results for an open system with on-ramp are presented in
Fig.~\ref{f.ramp}. We first examine the so-called {\it pinch
effect}~\cite{3PhTr,EmpR,KeCL} describing the process of a local
self-compression in synchronized regions, which leads to the
formation of small narrow jams. These small jams evolve finally
into a few wide jams through a merging process while moving
upstream when the small jams grow enough to lose the stability of
the synchronized flow. This process can be seen in
Fig.~\ref{f.ramp}(a). Empirically, such a process is the most
frequent type of congested traffic near a bottleneck and is thus
named the {\it general pattern} (GP)~\cite{EmpR}. This process is
captured more clearly in Fig.~\ref{f.ramp}(b). Starting from the
top, it is seen that synchronized flow (high flux and slow
velocity) is formed near the on-ramp. In the following two
pictures, many small drops are merged into small narrow jams along
the upstream. Finally, a few wide jams remain far away upstream
from the on-ramp. The other known~\cite{EmpR} types of congested
traffic near bottlenecks are {\it localized synchronized flow
patterns} (LSP), {\it widening synchronized flow patterns} (WSP),
and {\it moving synchronized flow patterns} (MSP). It is stressed
here that these patterns are also reproduced in the present model
as shown in Figs~\ref{f.ramp}(c,d,e). Furthermore, we remark that
the phase diagram for the congested patterns
[Fig.~\ref{f.ramp}(f)] is quite comparable to those
in~\cite{EmpR,KeCL}.

\begin{figure}
\includegraphics[width=8.0cm, height=6.0cm]{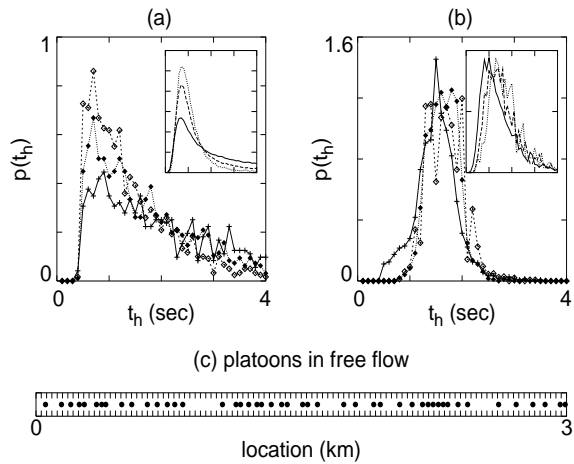}
\caption{(a) [(b)] Time-headway distribution of free
[synchronized] flow is depicted. The cross, filled diamond, and
open diamond represent densities of $12$, $16$, and $20$ veh/km
[$24$, $32$, and $40$ veh/km]. The insets are taken from
\cite{TiHe} (empirical data). Solid, dashed, and dotted lines
correspond to density ranges from $0\sim 12$, $12\sim 24$, and
$24\sim 36$ veh/km [$24\sim 36$, $36\sim 48$, and $48\sim 60$
veh/km]. (c) Randomly chosen part of the road whose overall data
give the connected open diamonds in (a).} \label{f.thd}
\end{figure}
Finally, we examine the time-headway distribution of the model.
The correspondence between the numerical results and its empirical
counterpart is quite satisfactory as shown in Fig.~\ref{f.thd}(a).
In contrast to the result in~\cite{KnBL}, the peaks below $1$ sec
are realized merely with identical vehicles. A typical spatial
configuration of the free flow shown in Fig.~\ref{f.thd}(c)
indicates that such peaks are attributed to the platoon formation
of vehicles ({\it platoon effect}). Revisiting Eqs.~(\ref{dcf}),
(\ref{gamma}), and (\ref{modtd}) with the requirement of free flow
($c_n^{t+1} = v_{n+1}^t = v_{\rm max}$ for all $n$) helps us to
understand the formation of such platoons. Then Eq.~(\ref{dcf})
allows even $0.3$ sec as a time-headway, which opens up the
possibility of the formation of vehicular platoons in free flow.
It is emphasized here that these platoons can explain the small
time-headway frequently observed in free flow without an
unrealistic high flux. Thus the platoons may be one of the
fundamental objects which can characterize free flow. Another
important role of the platoons is suggested as follows. Since the
large gaps between the platoons can absorb fluctuations
propagating backwards, they stabilize free flow and thus influence
the stability and transition properties. For the case of
synchronized flow, qualitative agreement with empirical
findings~\cite{TiHe} is observed as follows [see
Fig.~\ref{f.thd}(b)]: i) distribution shifts to the left as
density increases, ii) peaks are near $1.5$ sec, iii) time-headway
smaller than $1$ sec still exists, and iv) large time-headway
events are reduced compared to those of free flows.

In conclusion, a new CA traffic model focusing on the mechanical
restriction realized by limited acceleration and braking
capabilities is introduced. A further element, namely {\it human
overreaction}, is implemented in order to reflect the driver's
tendency toward biased reaction according to the local traffic
conditions. It is shown that the model reproduces most empirical
findings including the three known traffic phases, the so-called
{\it pinch effect} and several types of congested traffic patterns
as well as the small time-headway below 1 sec especially in free
flow. In particular, the presented model produces vehicular
platoons as a non-trivial element of free flow. Thus the {\it
platoon effect} is proposed to be the origin of many features of
free flow. We remark that some situations requiring greater
deceleration beyond the braking capacity for safety, such as
careless insertion near on-ramp, can result in collisions in this
model. In summary, it can be stressed that the presented model
sheds light on open questions of traffic modelling and is
therefore useful for the understanding of certain traffic
phenomena as well as for applications such as fully automated
driving.

H. K. L. was supported by the Brain Korea 21 Program. The authors
are grateful to the BMBF within the project DAISY for financial
support.


\begin{thebibliography}{99}

\bibitem{tgf03kerner} B. S. Kerner, in {\it Traffic and Granular Flow `01},
        edited by M. Fukui, Y. Sugiyama, M. Schreckenberg, and D. Wolf,
        Springer, Heidelberg, 13, (2003)
\bibitem{HeRe} D. Helbing,
        Rev. Mod. Phys. {\bf 73}, 1067 (2001)
\bibitem{ChRe} D. Chowdhury, L. Santen, and A. Schadschneider,
        Phys. Rep. {\bf 329}, 199 (2000)
\bibitem{Hys} I. Treiterer and J. A. Myers, in Proceedings of the 6th
        International Symposium on Transportation and Traffic Theory,
        edited by D. J. Buckley Elsevier, New York, (1974)
\bibitem{3Ph} B. S. Kerner and H. Rehborn,
        Phys. Rev. E {\bf 53}, R4275 (1996)
\bibitem{3PhTr} B. S. Kerner,
        Phys. Rev. Lett {\bf 81}, 3797 (1998)
\bibitem{EmpR} B. S. Kerner,
        Phys. Rev. E {\bf 65}, 046138 (2002)
\bibitem{TiHe} W. Knospe, L. Santen, A. Schadschneider, and M. Schreckenberg,
        Phys. Rev. E {\bf 65}, 056133 (2002)
\bibitem{NSCA} K. Nagel and M. Schreckenberg,
        J. Physique I {\bf 2}, 2221 (1992)
\bibitem{BaOp} M. Bando, K. Hasebe, A. Nakayama, A. Shibata, and Y. Sugiyama,
        Phys. Rev. E {\bf 51}, 1035 (1995)
\bibitem{KrLd} S. Krauss, P. Wagner, and C. Gawron,
        Phys. Rev. E {\bf 55}, 5597 (1997)
\bibitem{KeFl} B. S. Kerner and P. Konh\"{a}user,
        Phys. Rev. E {\bf 48}, R2335 (1993)
\bibitem{BeLi} P. Berg, A. Mason, and A. Woods,
        Phys. Rev. E {\bf 61}, 1056 (2000)
\bibitem{HKLi} H. K. Lee, H.-W. Lee, and D. Kim,
        Phys. Rev. E {\bf 64}, 056126 (2001)
\bibitem{HeLi} D. Helbing, A. Hennecke, V. Shvetsov, and M. Treiber,
        Math. Comput. Model. {\bf 35}, 517 (2002)
\bibitem{HKLi2} H. K. Lee, H.-W. Lee, and D. Kim,
        Phys. Rev. E {\bf 69}, 016118 (2004)
\bibitem{KnBL} W. Knospe, L. Santen, A. Schadschneider, and M. Schreckenberg,
        J. Phys. A {\bf 33}, L477 (2000)
\bibitem{KeCL} B. S. Kerner and S. L. Klenov,
        J. Phys. A: Math. Gen. {\bf 35}, L31 (2002);
        B. S. Kerner, S. L. Klenov, and D. E. Wolf,
        J. Phys. A: Math. Gen. {\bf 35}, 9971 (2002)
\bibitem{NoCo} N. Eissfeldt and P. Wagner,
        Eur. Phys. J. B {\bf 33}, 121 (2003)
\bibitem{Barlo1} R. Barlovic, L. Santen, A. Schadschneider, and M. Schreckenberg,
        Eur. Phys. J. B. {\bf 5}, 793, (1998)
\bibitem{Barlo2} R. Barlovic, T. Huisinga, A. Schadschneider, and M. Schreckenberg,
        Phys. Rev. E {\bf 66}, 046113 (2002)
\bibitem{Comment1} Limited deceleration has been examined similarly in
Ref.~\cite{KrLd} for a space continous model. However, that work
did not take account of the possibility that its safety velocity
(the counterpart of $c_n^{t+1}$) can not be realizable with
limited deceleration.

\end{thebibliography}
\end{document}